\documentclass[%
 reprint,
superscriptaddress,
 amsmath,amssymb,
 aps,
]{revtex4-2}
\usepackage{comment}
\usepackage{graphicx}% Include figure files
\usepackage{dcolumn}% Align table columns on decimal point
\usepackage{bm}% bold math
\usepackage{xcolor}
\usepackage{units}
\usepackage{appendix}

\usepackage[version=4]{mhchem}
\usepackage{tabularx}
\usepackage{multirow}
\usepackage{soul}
\usepackage[colorlinks = true, linkcolor = blue, urlcolor  = blue, citecolor = blue, anchorcolor = blue]{hyperref}

\begin{document}

\title{Heat-conserving three-temperature model for ultrafast demagnetization of $3d$ ferromagnets}% 

\author{M. Pankratova}
\email{maryna.pankratova@physics.uu.se}
\affiliation{Department of Physics and Astronomy, Uppsala University, Box 516, SE-75120 Uppsala, Sweden}

\author{I. P. Miranda}
\affiliation{Department of Physics and Astronomy, Uppsala University, Box 516, SE-75120 Uppsala, Sweden}

\author{D. Thonig}
\affiliation{School of Science and Technology,  \"Orebro University, SE-701 82, \"Orebro, Sweden}
\affiliation{Department of Physics and Astronomy, Uppsala University, Box 516, SE-75120 Uppsala, Sweden}

\author{M. Pereiro}
\affiliation{Department of Physics and Astronomy, Uppsala University, Box 516, SE-75120 Uppsala, Sweden}

\author{E. Sjöqvist}
\affiliation{Department of Physics and Astronomy, Uppsala University, Box 516, SE-75120 Uppsala, Sweden}

\author{A. Delin}
\affiliation{Department of Applied Physics, School of Engineering Sciences, KTH Royal Institute of Technology, AlbaNova University Center, SE-10691 Stockholm, Sweden}
\affiliation{SeRC (Swedish e-Science Research Center), KTH Royal Institute of Technology, SE-10044 Stockholm, Sweden}

\author{P. Scheid}
\affiliation{Universit\'e de Lorraine, LPCT, CNRS, UMR 7019, BP 70239, 54506 Vandoeuvre-l\'{e}s-Nancy Cedex, France}
\affiliation{Universit\'e de Lorraine, IJL, CNRS, UMR 7198, BP 70239, 54000 Nancy Cedex, France}
\affiliation{Department of Physics and Astronomy, Uppsala University, Box 516, SE-75120 Uppsala, Sweden}

\author{O. Eriksson}
\affiliation{Department of Physics and Astronomy, Uppsala University, Box 516, SE-75120 Uppsala, Sweden}

\author{A. Bergman}
\affiliation{Department of Physics and Astronomy, Uppsala University, Box 516, SE-75120 Uppsala, Sweden}

\date{\today}

\begin{abstract}
We study the ultrafast magnetization dynamics of bcc Fe and fcc Co using the recently suggested heat-conserving three-temperature model (HC3TM), together with atomistic spin- and lattice dynamics simulations. It is shown that this type of Langevin-based simulation is able to reproduce observed trends of the ultrafast magnetization dynamics of fcc Co and bcc Fe, in agreement with previous findings for fcc Ni. The simulations are performed by using parameters that to as large extent as possible are obtained from electronic structure theory. The one parameter that was not calculated in this way, was the damping term used for the lattice dynamics simulations, and here a range of parameters were investigated. It is found that this term has a large influence on the details of the magnetization dynamics.
The dynamics of iron and cobalt is compared with previous results for nickel and similarities and differences in the materials' behavior are analysed following the absorption of a femtosecond laser pulse. Importantly, for all elements investigated so far with this model, we obtain a linear relationship between the value of the maximally demagnetized state and the fluence of the laser pulse, which is in agreement with experiments.
\end{abstract}

\maketitle

\section{INTRODUCTION}

Ultrafast demagnetization was discovered by Beaurepaire and coauthors in 1996 \cite{beaurepaire1996ultrafast}. They observed demagnetization in a nickel film on picosecond timescales following the absorption of a femtosecond laser pulse. From the point of view of applications, ultrafast demagnetization is an important process in all-optical magnetization switching as well as for new applications in magnetic data storage and spintronics \cite{RevModPhys.82.2731}. 
In the same pioneering work of Beuarepaire \cite{beaurepaire1996ultrafast}, these experimental observations were interpreted using a three-temperature model (3TM) \cite{beaurepaire1996ultrafast}, which assumes three thermalised reservoirs, in particular, spin-, lattice, and electron reservoirs, that can exchange energy through coupling parameters (via electron-phonon, electron-spin, and spin-lattice coupling). The 3TM is often used to interpret ultrafast magnetization dynamics processes \cite{beaurepaire1996ultrafast,SCHEID2022169596,PhysRevB.106.174407}. Recently, many other models have been proposed \cite{RevModPhys.82.2731,SCHEID2022169596,PhysRevB.92.064403,shim2020role} to describe possible mechanisms of ultrafast demagnetization, such as the importance of spin-dependent transport of laser-excited electrons \cite{PhysRevLett.105.027203}, the optical inter-site spin transfer effect (OISTR) that was considered in Ref.\,\cite{willems2020optical}, and the Elliott–Yafet electron-phonon spin-flip scattering, studied in Ref.\,\cite{PhysRevB.78.174422}. In \cite{PhysRevB.85.184301} a dynamic spin-lattice-electron model  was proposed. Using this model, the authors calculated laser-induced demagnetization of iron thin film and obtained very accurate agreement with the experimental observations. One of the important outcomes of the \cite{PhysRevB.85.184301} is also establishing the relation between the dissipative parameters entering the Langevin equations for the lattice and spin degrees of freedom and the heat transfer coefficients of 3TM.

In addition, Zahn \textit{et al.} \cite{PhysRevResearch.3.023032,PhysRevResearch.4.013104} proposed an energy-conserving model based on atomistic spin dynamics simulations. Using this model they studied  ultrafast demagnetization of nickel \cite{PhysRevResearch.3.023032}, iron and cobalt \cite{PhysRevResearch.4.013104}. However, the model of Zahn \textit{et al.} relies on the microscopic electron-lattice heat transfer coefficient which is hard to estimate. Literature values for the electron-phonon coupling parameter G$_{ep}$ can vary for nickel up to one order of magnitude \cite{PhysRevResearch.3.023032, PhysRevB.106.174407}. For iron, reported values of the electron-phonon coupling parameter differs from $7 \cdot 10^{17}$ to $\unit[5.48 \cdot 10^{18}]{W/ m^3 K}$  \cite{PhysRevResearch.4.013104}, and for cobalt from $6 \cdot 10^{17}$ to $\unit[4.05 \cdot 10^{18}]{W/ m^3 K}$ (please see \cite{PhysRevResearch.4.013104} and references therein).

One of the reasons for this uncertainty of the electron-lattice heat transfer coefficient is that the electron-phonon coupling is often deduced from observables, without taking into account the energy cost of demagnetization. Also, these parameters are often estimated indirectly using, for example, a two-temperature model \cite{PhysRevX.6.021003}. It was demonstrated that $G_{ep}$ extracted from the data employing the two-temperature model is twice smaller than that calculated from ab initio theory \cite{PhysRevX.6.021003}. 
These differences in reported $G_{ep}$ values makes the interpretation
of the experimental observations very challenging. 

To address this, a previous work proposed a heat-conserving three-temperature model (HC3TM) \cite{PhysRevB.106.174407} for calculation on spin, lattice and electron temperatures during atomistic spin-dynamics simulations. The benefit of this model is that it reduces the dependence on heat-transfer parameters that are hard to estimate, such as electron-phonon, $G_{ep}$, electron-spin, $G_{es}$, and the spin-lattice coupling, $G_{sl}$. In addition, this model allows for a description of the nickel demagnetization rate on sub-picosecond timescales. 

In the present work, we apply the HC3TM model to study ultrafast demagnetization of bcc Fe and fcc Co and compare with previously published data on ultrafast dynamics of fcc Ni \cite{PhysRevB.106.174407}. The ultrafast demagnetization of these three ferromagnets was experimentally studied in Ref.\,\cite{scheid2023uncovering}. It was observed that the amplitude of the demagnetization value increases linearly with the fluence of the light for Fe, Ni, and Co. A model to explain the experimental observations was proposed based on the assumption that the linear dependence of demagnetization on fluence is driven by the increase in temperature, the electron-phonon coupling, the electron-magnon scattering, and a reduction of the inter-atomic exchange. In the present work, we aim to study the impact of fluence on the demagnetization of 3d ferromagnets using the HC3TM. Moreover, we investigate in more detail the impact of various factors on ultrafast demagnetization, such as Gilbert damping, lattice damping, microscopic spin-lattice coupling, and laser fluence. 

\section{Atomistic spin-lattice dynamics simulations}

Spin and lattice dynamics is obtained from coupled atomistic spin- and lattice dynamics (SLD) simulations, using Langevin dynamics simulations, as proposed in Ref.\,\cite{PhysRevB.99.104302}:
\begin{equation}
\label{eq1}
\begin{split}
    \frac{d \boldsymbol{m}_i}{d t}= - \frac{\gamma}{(1+\alpha ^2)}\boldsymbol{m}_i\times(\boldsymbol{B}_i+\boldsymbol{B}_i^{fl})\\
    - \frac{\gamma}{(1+\alpha ^2)}\frac{\alpha}{m_i}\boldsymbol{m}_i\times(\boldsymbol{m}_i \times [\boldsymbol{B}_i+\boldsymbol{B}_i^{fl}])
    \end{split}
\end{equation}
\begin{equation}
\label{eq2}
    \frac{d \boldsymbol{u}_k}{dt} = \boldsymbol{v}_k
\end{equation}
\begin{equation}
\label{eq3}
    \frac{d \boldsymbol{v}_k}{dt}= \frac{\boldsymbol{F}_k}{M_k}+\frac{\boldsymbol{F}_k^{fl}}{M_k} - \nu \boldsymbol{v}_k,
\end{equation}
where $\boldsymbol{m_i}$ represents an atomic magnetic moment, $m_i$ and  $\gamma$ are the saturation magnetization and the gyromagnetic ratio correspondingly. Atomic displacements are denoted by $\boldsymbol{u_k}$, and velocities are $\boldsymbol{v_k}$. We obtain an effective exchange field $\boldsymbol{B}_i = - \partial H_{\mathrm{SLD}}/\partial \boldsymbol{m}_i$ from the spin-lattice Hamiltonian, $H_{SLD}$. The Hamiltonian used in this work includes magnetic, lattice and spin-lattice coupling parts following Ref.\,\cite{PhysRevB.99.104302}: 
\begin{equation}
    H_{\mathrm{SLD}}= H_{\mathrm{LL}}+H_{\mathrm{SS}}+H_{\mathrm{SSL}}
\end{equation}
Concretely:

\begin{equation}
    H_{\mathrm{LL}} = \frac{1}{2} \sum_{kl} \Phi_{kl}^{\mu \nu} u_k^{\mu} u_l^{\nu} + \frac{1}{2} \sum_{k} M_{k} \nu_k^{\mu} \nu_k^{\mu}
\end{equation}
where the force constant tensor $\Phi_{kl}^{\mu \nu}$ is a rank 2 tensor in real space, and $M_k$ is the mass of atom $k$.  The magnetic Hamiltonian is described by

\begin{equation}
    H_{\mathrm{SS}} = - \frac{1}{2} \sum_{ij} J_{ij}^{\alpha \beta}(0) m_i^{\alpha} m_j^{\beta}
\end{equation}
where $J_{ij}^{\alpha \beta}(0)$ is the exchange tensor at the equilibrium lattice positions. $\alpha,\beta \in 
\{ x,y,z\}$ denote Cartesian components in spin space, while $\mu,\nu \in 
\{ x,y,z\}$ corresponds to Cartesian components in real space.

It was recently suggested that spin-lattice coupling can be represented in coupled spin-lattice simulations, by a dependence of the exchange interaction on atomic displacements $\boldsymbol{u_k}$, resulting in functions of the type $J_{ij}^{\alpha \beta}(\boldsymbol{u_k})$ (see Ref.\,\cite{PhysRevB.99.104302}). As the values of $\boldsymbol{u_k}$ are usually small, the spin-lattice term in the Hamiltonian can be obtained by Taylor expanding the magnetic bi-linear Hamiltonian with respect to the lattice displacements. 
This results in a spin-lattice coupling term that is bi-linear in spin and linear in displacements \cite{PhysRevB.99.104302}. Concretely,
\begin{equation}
    H_{\mathrm{SSL}} = - \frac{1}{2}\sum_{ijk} \Gamma_{ijk}^{\alpha \beta \mu} m_i^{\alpha} m_j^{\beta} u_k^{\mu},
    \label{SLHAM}
\end{equation}
where we introduce the coupling constant $\Gamma_{ijk}^{\alpha\beta\mu}\equiv\Gamma_{ijk}^{\alpha\beta\mu}(u_k^{\mu}) =\partial J_{ij}^{\alpha \beta}(u_k^{\mu})/\partial u_k^{\mu}$. 

All parameters for the spin-lattice dynamics simulations were obtained from \textit{ab-initio} calculations (see Appendix \ref{AppendixB} for  details). This includes exchange interactions, magnetic moments, inter atomic forces, and spin-lattice couplings. The force at site $k$ is defined by $\boldsymbol{F}_k = - \partial H_{\mathrm{SLD}}/\partial \boldsymbol{u}_k$.  Gilbert and lattice damping constants are denoted $\alpha$ and
$\nu$, respectively. In these types of Langevin simulations one employs stochastic fields, $\boldsymbol{B}_i^{fl}$ and $\boldsymbol{F}_{k}^{fl}$, as white noise with properties $\langle B_{i,\mu}^{fl}(t) B_{j,\nu}^{fl}(t') \rangle=2D_M \delta_{ij}\delta_{\mu\nu}\delta(t-t')$ and $\langle F_{i,\mu}^{fl}(t) B_{j,\nu}^{fl}(t') \rangle=2D_L \delta_{kl}\delta_{\mu\nu}\delta(t-t')$. In our simulations, we use $D_M= \alpha k_B T/\gamma m$, $D_L= \nu M k_B T$ , where $T$ and $k_B$ are temperature and Boltzmann constant respectively (for details see e.g. Ref.\,\cite{ErikssonAtomistic}).

\section{Heat-conserving three-temperature model}
\label{HC3TM}

\begin{figure}[h]
\begin{tabular}{cc}
(a) & \includegraphics[width=0.33\textwidth]{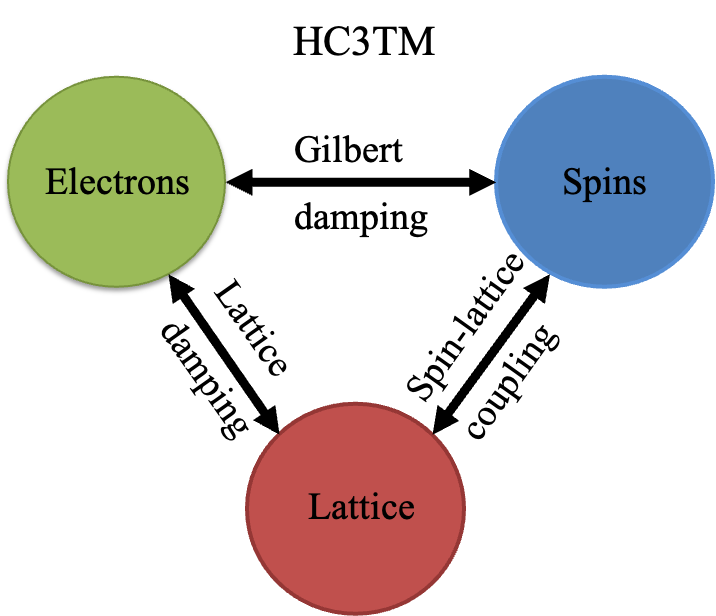} \\
(b) &\includegraphics[width=0.33\textwidth]{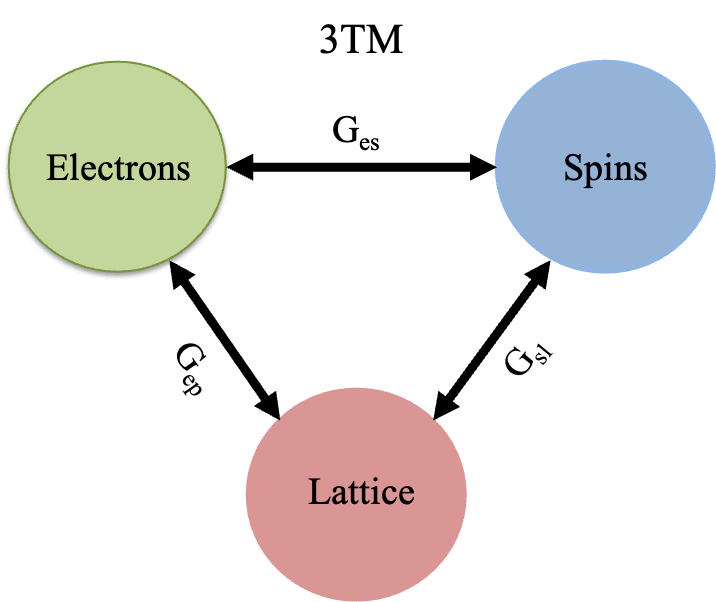}
\end{tabular}
\caption{\label{fig:TM_scheme} The main idea behind the HC3TM (a) and 3TM (b) models. The text along the arrows indicates the parameters responsible for heat transfer in each model. For the parameters of the 3TM, see Ref.\,\cite{beaurepaire1996ultrafast}.
}
\end{figure}

The heat-conserving three-temperature model was proposed in Ref.\,\cite{PhysRevB.106.174407} for the study of ultrafast demagnetisation of fcc Ni. Similar to Beurepaire's three-temperature model, it is assumed that lattice, and spin systems are connected to a finite electronic heat bath. Both spin and lattice subsystems have their own damping; Gilbert or lattice damping, correspondingly. The model does not explicitly rely on the heat transfer parameters of the conventional 3TM \cite{beaurepaire1996ultrafast}. The difference between HC3TM and the conventional 3TM is easy to see from the schematic picture in Fig.\,\ref{fig:TM_scheme}.  Overall, instead of microscopic heat transfer coefficients, as used in 3TM, the HC3TM is formulated in terms of atomistic damping coefficients. Thus, the HC3TM calculates the heat transfer between the three reservoirs (spin, lattice and electron)\cite{beaurepaire1996ultrafast} directly from the simulations. In fact, all parameters of this theory can be calculated using, for example, density functional theory, although we note that in the present work we varied the lattice damping parameter in a range of realistic values, as opposed to performing an explicit calculation of it. The relation between the lattice damping and $G_{ep}$ was established in \cite{PhysRevB.85.184301} using spin-lattice-electron three-temperature model. It was shown that the lattice damping is directly proportional to $G_{ep}$, which was previously also reported by Duffy et el. in \cite{duffy2006including,duffy2009including}.
Another distinction of HC3TM is that (unlike 3TM) the temperature of the different subsystems is calculated at every time step of the spin-lattice simulation using the expression:

\begin{equation}
    \Delta T_e (t) = -\frac{C_l(T_l)}{C_e(T_e)}T_l(t) - \frac{C_s(T_s)}{C_e(T_e)}T_s(t) + \frac{W(t)}{C_e(T_e)},
    \label{eq5}
\end{equation}
where $T_l$ is calculated from the average kinetic energy of the lattice vibrations; $\langle E^{kin}_l\rangle/k_B$. The spin temperature, $T_s$ is calculated according to \cite{PhysRevE.82.031111}, i.e. using $T_s=\frac{\langle\sum_{i}\left|\hat{\boldsymbol{m}}_{i}\times\boldsymbol{B}_i\right|^2\rangle}{2k_B\langle\sum_{i}\hat{\boldsymbol{m}}_{i}\cdot\boldsymbol{B}_i\rangle}$, where $\hat{\boldsymbol{m}_{i}}$ is the normalized local spin moment. The definition of an instantaneous measure of the temperatures in an out-of-equilibrium system is not obvious \cite{PhysRevE.49.1040}, but is still typically assumed to hold in these kinds of models. Introducing a time average of the temperatures, for narrow time windows, does however not change the results significantly.
In addition, Eq.\,\eqref{eq5} 
contains the temperature dependent specific heats of the electron-, lattice- and spin systems, $C_e(T)$, $C_l(T)$ and $C_s(T)$, respectively. The calculations details of the specific heat capacities are given in Appendix \ref{AppendixA}. The possibility for an external stimulus, e.g. a laser, to momentarily increase the temperature of the electronic subsystem is captured by the source term $W(t)$ of Eq.\,\eqref{eq5}, which is modelled as a Gaussian function: 
$W(t)=W_0 \exp(-(t-t_0)^2/2 \sigma^2)
$, where $t_0$ is the center, $\sigma = 0.02$ ps is the width of the pulse, and $W_0$ is the magnitude (in Joules), correspondingly. More details concerning the HC3TM and its comparison with 3TM can be found in Ref.\,\cite{PhysRevB.106.174407}.

\section{Results}
\subsection{HC3TM for cobalt and iron}

In this section we present results of spin-lattice dynamics simulations of fcc Co and bcc Fe. In both cases, we run simulations for cells with a $60 \times 60 \times 60$ repetition of the fcc (bcc) unit cell, using periodic boundary conditions. In addition, we used $N_t = 1 \times 10^6$ time steps of $dt = \unit[10^{-16}]{s}$, in combination with other parameters presented in Table \ref{tab:table1}. 

\begin{figure}[tbh]
\includegraphics[width=0.47\textwidth]{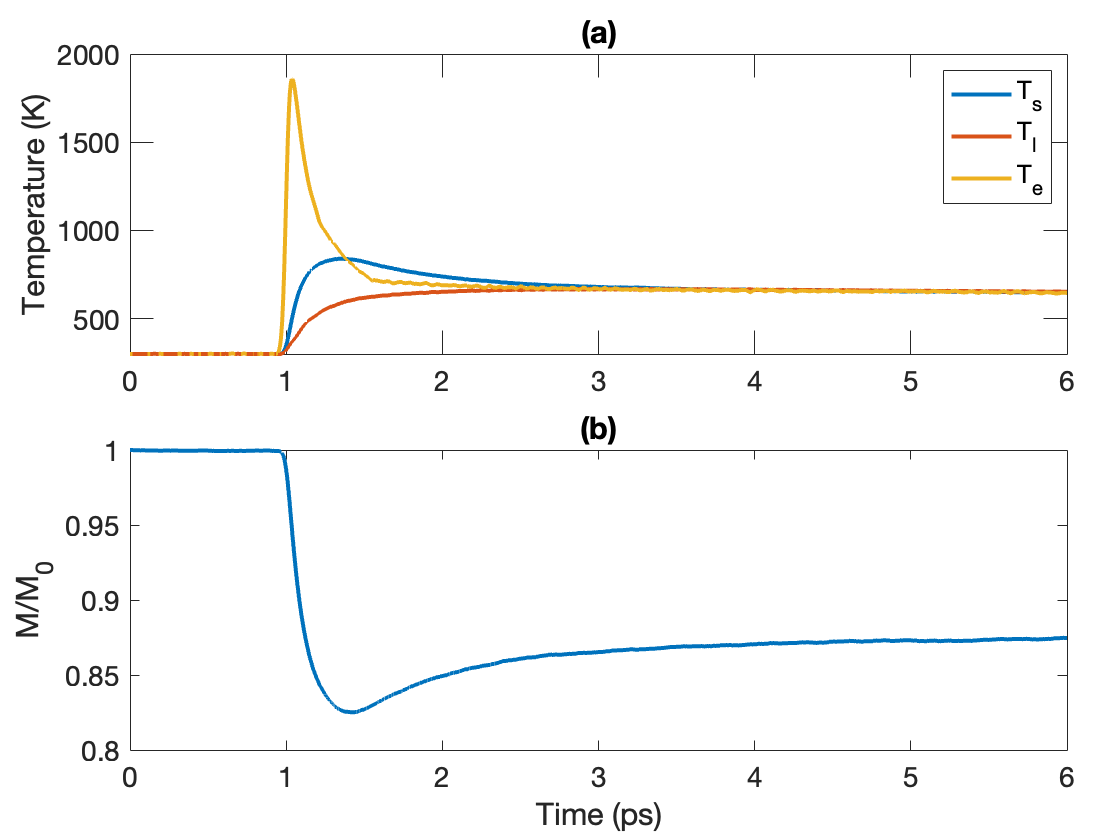}
\caption{\label{fig:magn_all_Co} Spin, lattice and electron temperatures (a) and magnetization dynamics (b) of fcc Co, obtained with the HC3TM model. The pulse fluence is 45 J/m$^2$.}
\end{figure}

We start by describing results from ultrafast demagnetization of fcc Co. First, we present temperature profiles (Fig.\,\ref{fig:magn_all_Co}) for spin, lattice, and electron subsystems, for a laser fluence of $\unit[45]{J/m^2}$, together with the corresponding magnetization dynamics. Other parameters relevant for these simulations are listed in Table \ref{tab:table1}. It can be seen from the figure that similar to experimental observations \cite{doi:10.1063/5.0049692} and theoretical studies \cite{PhysRevResearch.4.013104} demagnetization happens on subpicosecond timescales, followed by a remagnetization that has a faster recovery after $\sim$ 1 ps and a slower remagnetization after that. We obtain values of the position of the magnetization minima that are close to experimental values, and we also obtain demagnetization/remagnetization timescales that are consistent with experimental observations. It is important to note that in Ref.\,\cite{PhysRevResearch.4.013104}, using the parametrized three-temperature model, features of the magnetization dynamics similar to the data of Fig.\,\ref{fig:magn_all_Co} were also observed, however for an adjusted Gilbert damping value of $0.1$, which is unrealistically large. In contrast, in the calculations presented here we use a value of damping from ab-inito calculations (that is close to experiment), that as seen in Table \ref{tab:table1} is much smaller than the value used in Ref.\,\cite{PhysRevResearch.4.013104}. 

\begin{table}[t]
\caption{\label{tab:table1}%
List of the parameters used in the simulations.
}
\begin{ruledtabular}
\begin{tabular}{cr}
\textrm{Parameter}&
\textrm{Value}\\
\colrule
\\
\multicolumn{1}{l}{Iron}\\
\colrule
Gilbert damping $\alpha$ &  0.008 \cite{PhysRevB.85.184301}\\
Lattice damping $\nu$ & $1 \times 10^{-13}$ kg/s\\
\colrule
\\
\multicolumn{1}{l}{Cobalt}\\
\colrule
Gilbert damping $\alpha$ &  0.0024 (\cite{PhysRevB.95.134411}) or 0.0014 \cite{Lu2022} \\
Lattice damping $\nu$ & $1 \times 10^{-13}$ kg/s\\
\colrule
\\
\multicolumn{1}{l}{Nickel}\\
\colrule
Gilbert damping $\alpha$ &  0.024 (\cite{PhysRevB.106.174407})\\
Lattice damping $\nu$ & $1 \times 10^{-13}$ kg/s\\
\end{tabular}
\end{ruledtabular}
\end{table}

\begin{figure}[tbh]
\includegraphics[width=0.47\textwidth]{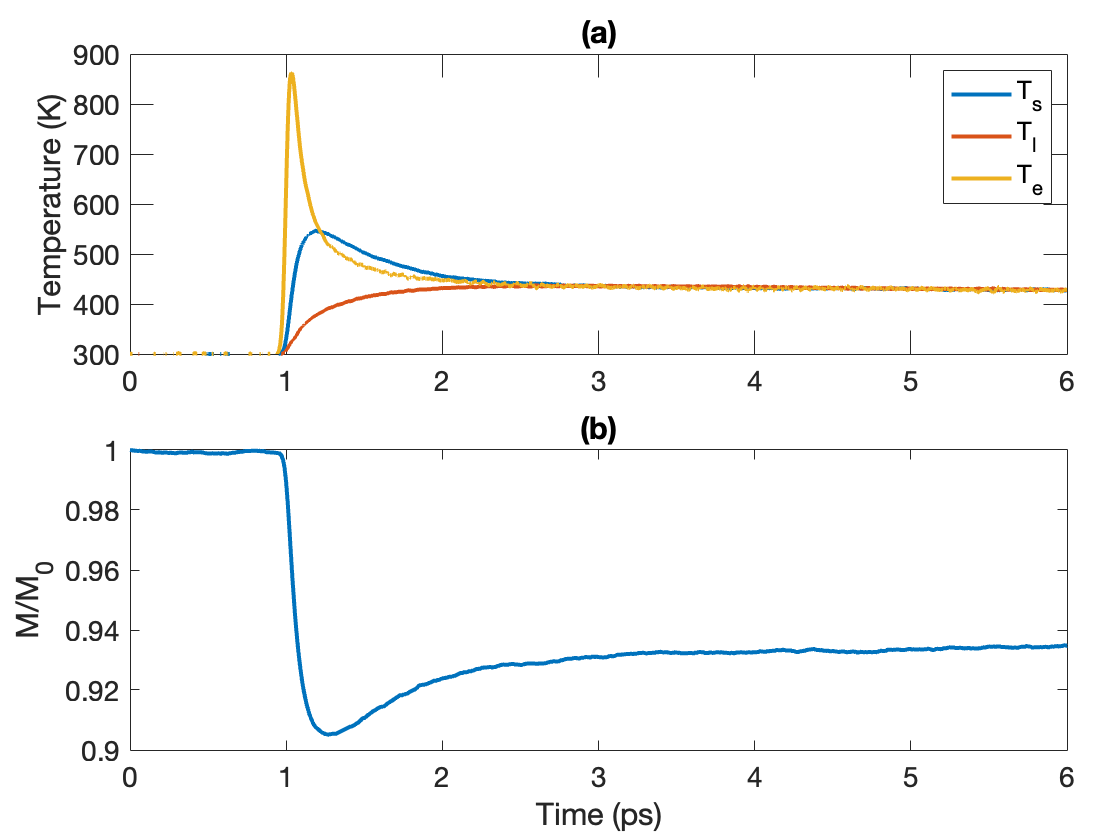}
\caption{\label{fig:magn_all_Fe} Spin, lattice and electron temperatures (a) and magnetization dynamics (b) of bcc Fe, obtained with the HC3TM model, pulse fluence is 15 J/m$^2$.}
\end{figure}

Similar spin-lattice simulations using the HC3TM, were performed for bcc Fe. Just like for fcc Co we first present magnetization dynamics after the absorption of a laser pulse and the corresponding spin, lattice, and electronic temperatures, see Fig.\,\ref{fig:magn_all_Fe}. Using the HC3TM model we obtain realistic demagnetization/remagnetization times in comparison with experimental studies and in this case even the maximally reduced magnetization is quite close to the experimentally observed value \cite{PhysRevB.78.174422}, for the same pulse fluence ($M/M_0$ is equal around $0.9$ in our calculations). 

\subsection{Results from coupled spin-lattice simulations}

To study the impact of direct spin-lattice coupling on the magnetization dynamics, given by Eq.\,\eqref{SLHAM}, we consider explicitly the spin-lattice coupling, where we obtain the elements of the tensor, $\Gamma_{ijk}$ in the spin-lattice coupling, from calculations based on density functional theory (e.g. as described in Ref.\,\cite{PhysRevB.99.104302}). Calculated values of these parameters were used to study the impact of direct spin-lattice coupling for bcc Fe, and fcc Co.
This explicit, calculated spin-lattice coupling allows for a direct exchange of heat between the spin and lattice subsystems, outside of the channel provided by the HC3TM. Note that the HC3TM can be applied whether or not the direct spin-lattice term in Eq.\,\eqref{SLHAM} is included or neglected, and the results that follow were obtained from the HC3TM with Eq.\,\eqref{SLHAM} included explicitly in the simulations. 

\begin{figure}[htb]
\includegraphics[width=0.47\textwidth]{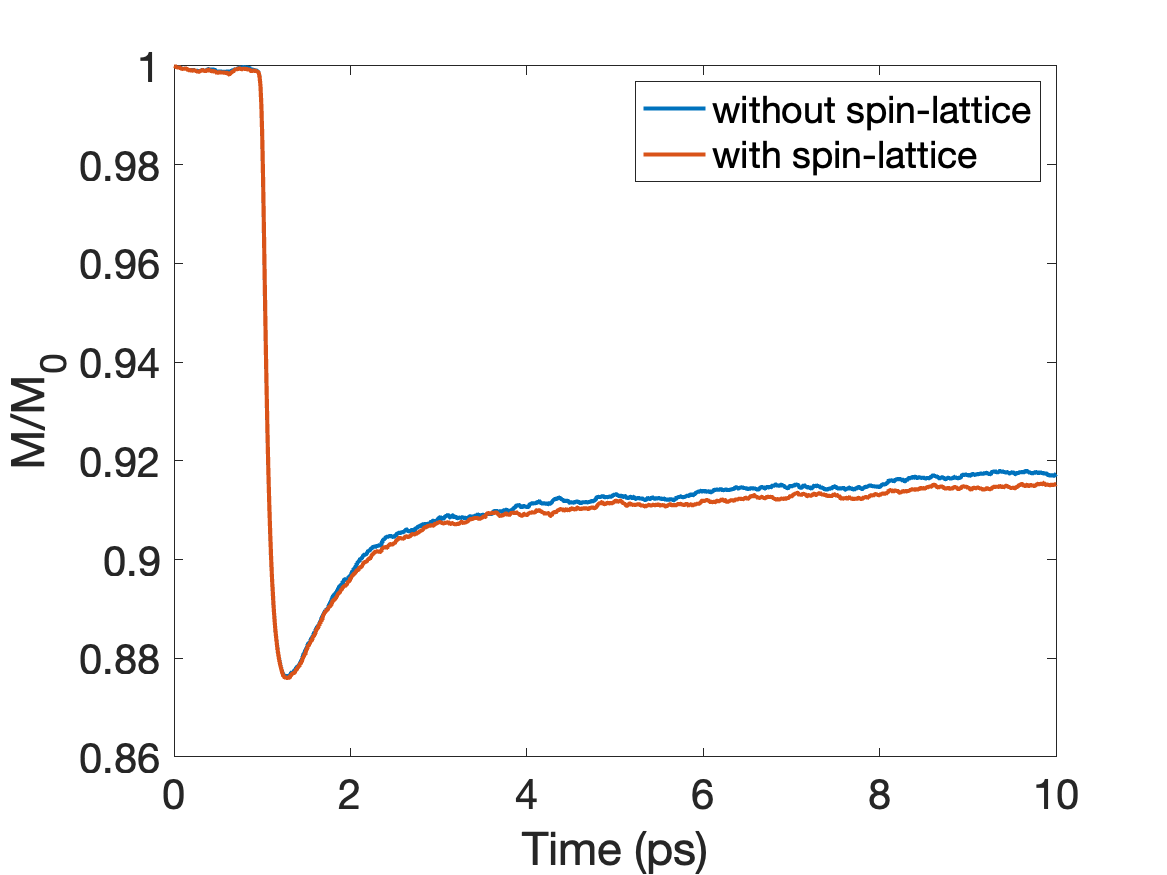}
\caption{\label{fig:magn_spin-latt} Impact of spin-lattice coupling on magnetization dynamics of bcc Fe. The pulse fluence is 20 J/m$^2$.}
\end{figure}

In Fig.\,\ref{fig:magn_spin-latt} we show for bcc Fe the demagnetization profile from coupled spin-lattice dynamics  simulations in combination with the HC3TM. The simulations are, as noted, done with and without the coupling term in Eq.\,\eqref{SLHAM}.
It can be seen from Fig.\,\ref{fig:magn_spin-latt} that spin-lattice coupling impacts the magnetization dynamics very weakly for bcc Fe. If any effect can be observed, it would be a slightly slower remagnetization rate when the spin-lattice coupling is included. The difference compared to the scenario without the spin-lattice coupling is however almost of the same order of magnitude as the numerical noise of the simulations. We note that a simulation which completely ignores this term still has ability to transfer heat from the spin system to the lattice, via Eq. (\ref{eq5}). A possible explanation for the low influence of the spin-lattice couplings can be drawn as follows: even at the maximum spin temperature in the simulations ($T_s\lesssim600$ K for Fe), the spin system is far away from $T_c$, i.e. with a general strong coupling between spin moments. Thus, the small average displacements (in the order of $10^{-2}a$) and their distribution throughout the sample result in the summed 
effect of Eq. \eqref{SLHAM} being only a minor perturbation on the total exchange field. 

Also, for bcc Fe the resulting profile is in rather good agreement with experimental observations. For materials with larger spin-lattice coupling\,\cite{PhysRevB.105.104418} it can be expected that the importance of Eq.\,\eqref{SLHAM} increases, and our simulations show that for bcc Fe the
spin-lattice coupling becomes more important for higher pulse fluences (data not shown). 

\begin{figure}[htb]
\includegraphics[width=0.48\textwidth]{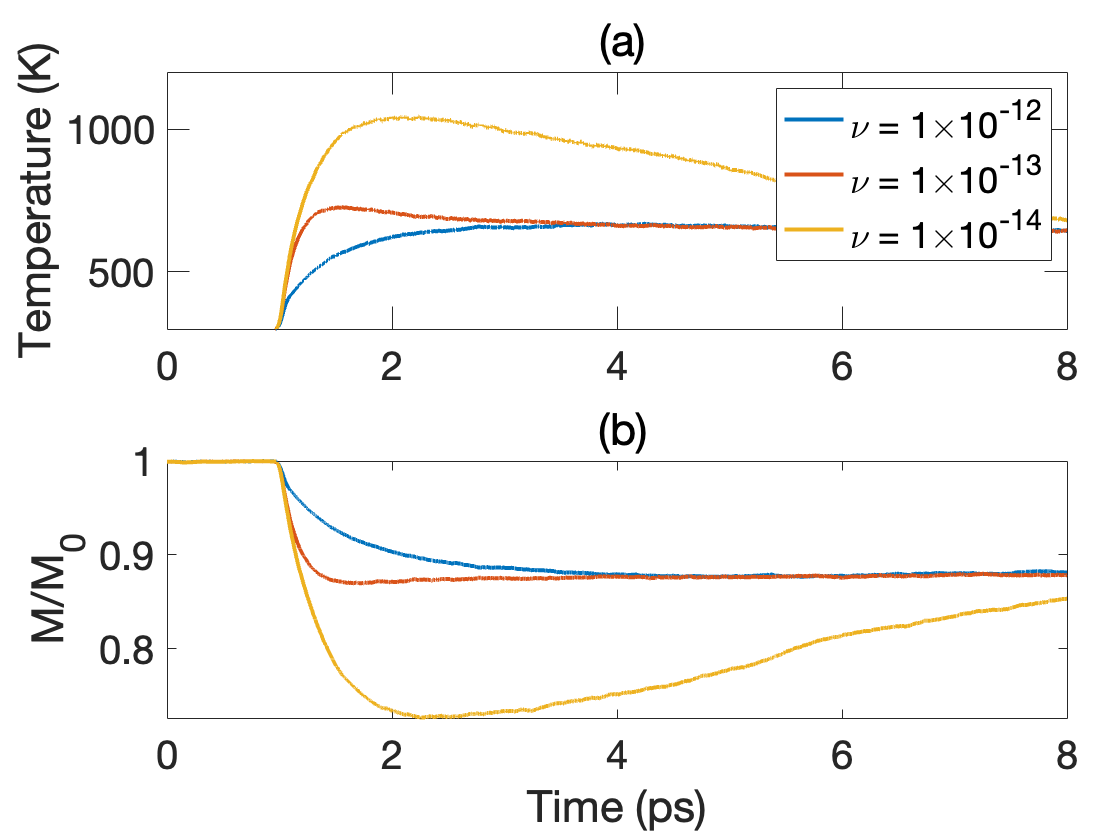}
\caption{
Spin temperature (a) and magnetization dynamics (b) of fcc Co for various lattice damping values used in HC3TM ($\nu$ in Eq.\,\eqref{eq3}), with temperature-dependent heat capacities (see Ref.\,\cite{PhysRevB.106.174407} for details). The pulse fluence is 45 J/m$^2$.}
\label{fig:magn_latt_Co} 
\end{figure}

While the impact of spin-lattice coupling, via Eq.\,\eqref{SLHAM}, on the magnetization dynamics is only marginal for the elemental ferromagnts, the impact of the dynamical properties of the lattice itself, and in particular, the lattice damping, is significant. We illustrate this for fcc Co, in Fig.\,\ref{fig:magn_latt_Co}. Here it is seen that a decrease of the lattice damping ($\nu$ in Eq.\,\eqref{eq3} leads to an increase of the demagnetization; both the magnetization minimum as well as the reduced magnetization ($M\over M_0$) are influenced by the lattice damping parameter. This can  best be understood from the fact that the coupled spin- and lattice system allows for heat to be dissipated from both reservoirs, when the lattice (and spin) damping parameter is larger. This causes the temperature profile to reach lower values and to equilibrate quicker both for the spin- and lattice system, as shown in  Fig.\,\ref{fig:magn_latt_Co}, with a resulting stronger impact on the magnetization profile for lower values of the lattice damping. 
This is detailed clearly when comparing the spin temperatures (Fig.\,\ref{fig:magn_latt_Co}) for various values of lattice damping. Lower lattice damping values lead to higher spin temperature and therefore, to larger magnetization drop and a longer remagnetization time. One of the important conclusions from this observation is that it is essential to take the lattice dynamics properly into account while studying ultrafast demagnetization. If the lattice is not considered at all, then one needs to note, while comparing simulations with experimental data, that this will lead to an overestimation of the magnetization drop and to longer remagnetization times, for a given laser pulse.

\subsection{Comparison of magnetization dynamics: bcc Fe, fcc Co, fcc Ni}

\begin{figure}[htb]
\includegraphics[width=0.47\textwidth]{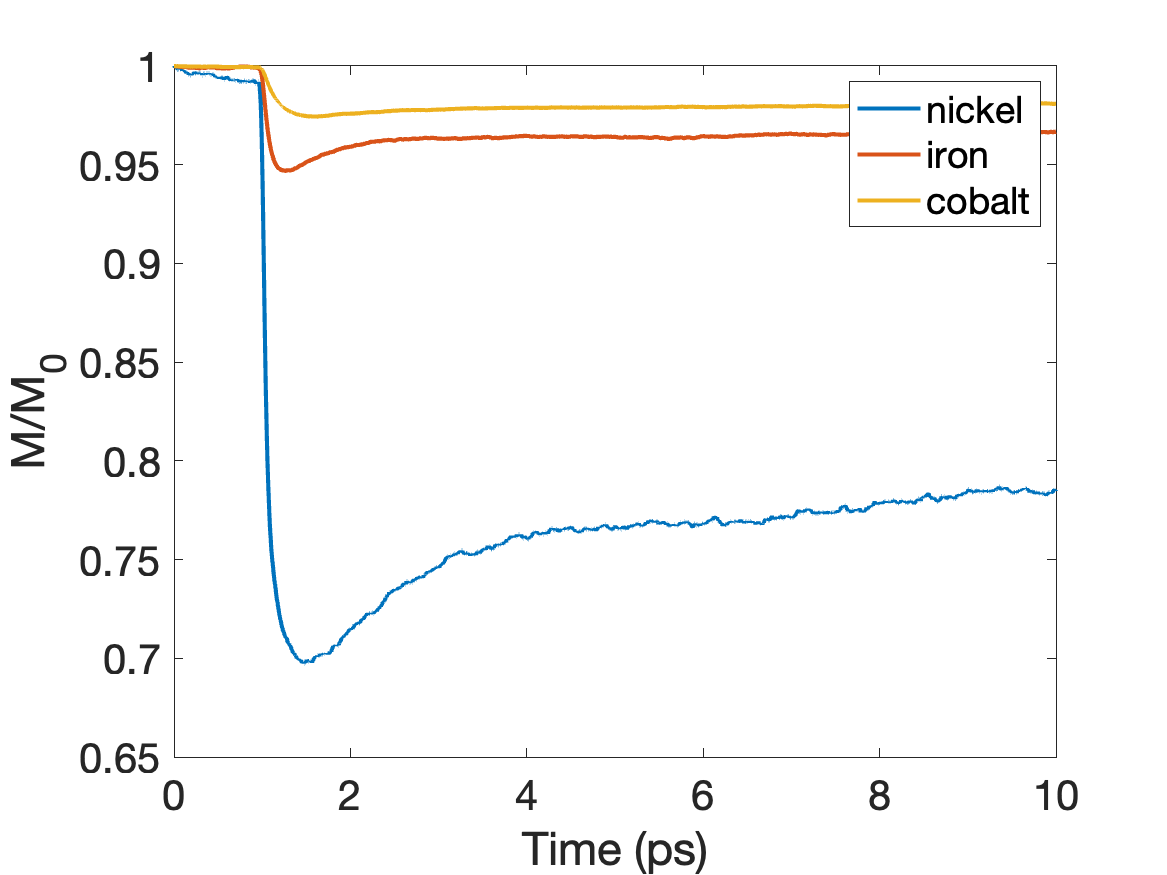}
\caption{Ultrafast demagnetization in fcc Ni, fcc Co and bcc Fe for the same value of the laser pulse fluence $\unit[8]{J/m^2}$.}
\label{fig:compare_all_same}
\end{figure}

\begin{figure}[htb]
\includegraphics[width=0.47\textwidth]{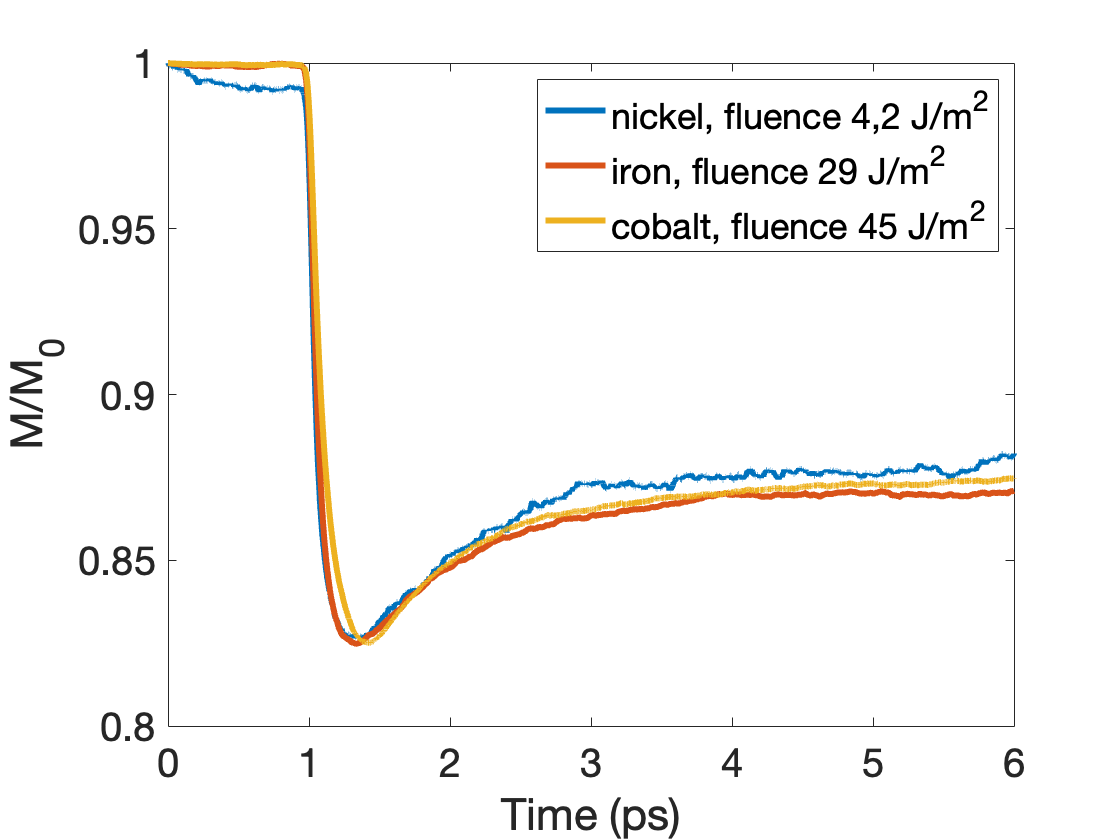}
\caption{ Ultrafast demagnetization in fcc Ni, fcc Co and bcc Fe for a different values of laser pulse fluence. The fluence is chosen to obtain the same demagnetization amplitude. 
}
\label{fig:compare_all_diff}
\end{figure}

In this section we discuss magnetization dynamics of bcc Fe and fcc Co. We compare the results from the two elements to previous results reported for fcc Ni, which also has been studied using the HC3TM \cite{PhysRevB.106.174407}. The results are summarized in Figs.\,\ref{fig:compare_all_same} and \ref{fig:compare_all_diff}.
We begin with comparing ultrafast magnetization dynamics of fcc Co, fcc Ni, and bcc Fe, for the same laser pulse fluence, presented in Fig.\,\ref{fig:compare_all_same}. It can be seen that in this case, the demagnetization for fcc Ni is the most prominent among the three systems, while the smallest effect is found for fcc Co. This difference in demagnetization properties is consistent with the difference in observed values of $T_c$, which is highest for fcc Co and smallest for fcc Ni.

The next step in our comparison is to choose values of the pulse fluence for all three materials resulting in the same (or very similar) magnetization dynamics.
 As one can see in Fig.\,\ref{fig:compare_all_diff}, due to very different $T_c$ values, we need a pulse fluence as high as $\unit[45]{J/m^2}$ for cobalt to exhibit a demagnetization that is similar to nickel for a fluence of $\unit[4.2]{J/m^2}$. It can also be seen from Fig.\,\ref{fig:compare_all_diff} that in this case the magnetization dynamics of all three systems are very similar. The remaining differences are minute, for example, the minima of cobalt's magnetization curve appears somewhat later than the one of iron and nickel. 

It is important to take into account that this very similar dynamics is observed for specific parameter values, such as the Gilbert and lattice damping. For example, to obtain the curve for cobalt presented in Fig.\,\ref{fig:compare_all_diff}, we used the  damping value from Ref.\,\cite{PhysRevB.95.134411}, see Table \ref{tab:table1}. However, with the damping value obtained from ab initio calculations, the magnetization dynamics will be somewhat different. In particular, we then observe much slower remagnetization of cobalt (as, for example, shown in Fig.\,\ref{fig:magn_latt_Co}  obtained with the Gilbert damping set to 0.0014).   

\begin{figure}[htb]
\includegraphics[width=0.47\textwidth]{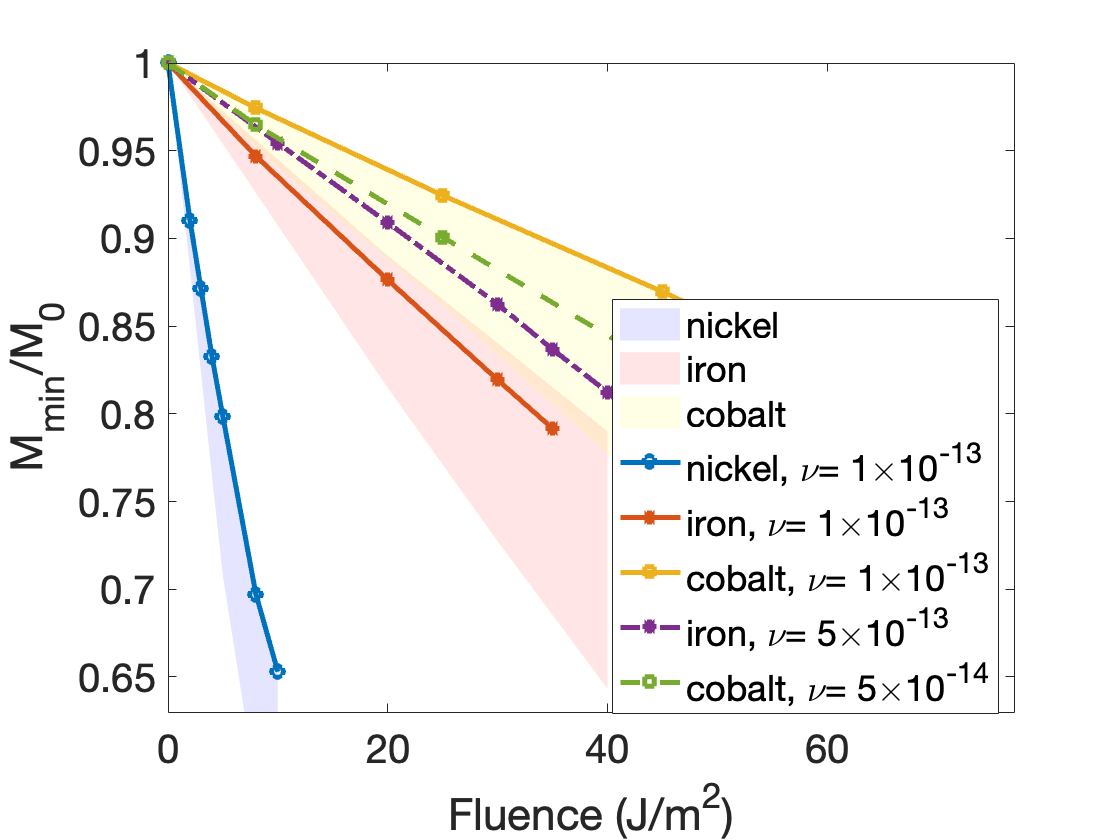}
\caption{\label{fig:magn_fluence} The dependence of the maximally demagnetized value versus laser fluence for bcc Fe, fcc Co and fcc Ni. The shaded area demonstrate a range of reported \cite{PhysRevResearch.3.023032, PhysRevB.106.174407, PhysRevResearch.4.013104} lattice damping parameters obtained using the relation between lattice damping and $G_{ep}$ from \cite{PhysRevB.85.184301}.}
\end{figure}

In Fig. \ref{fig:magn_fluence} we plot the maximally demagnetized state (minimum value of the magnetization curve, $M_{min}/M_0$) as function of absorbed laser pulse fluence for bcc Fe, fcc Co, and fcc Ni for the same value of the lattice damping (solid lines).  The first observation is that one needs the lowest laser fluence to demagnetize fcc Ni and the strongest for fcc Co. This can, as mentioned earlier in this paper, be connected to the corresponding $T_c$.  Secondly, $M_{min}/M_0$ in the simulations depends linearly on laser fluence, something which was observed experimentally in Ref.\,\cite{scheid2023uncovering} and also for other materials in Refs.\,\cite{windsor2022exchange,mishra2021ultrafast}.

In our work we use Gilbert damping that is the highest for nickel and the lowest for cobalt, which might add to the observed trend in Fig.\,\ref{fig:magn_fluence}. In particular, a higher Gilbert damping, which in the HC3TM corresponds to the electron-magnon coupling, leads to a stronger heat transfer from the electronic to the spin subsystem, and therefore faster demagnetization, as in the case of nickel. 

In the literature, the reported Gilbert damping parameters of iron, nickel, and cobalt deviate between different experiments. For example, reported $\alpha$ values for iron are in the range of $0.0019-0.0072$ \cite{Lu2022}. Nickel does have a consistently larger Gilbert damping than the other two elements, stemming from the $1/m_{Ni}$ scaling and the peak in the minority spin channel at the Fermi level \cite{Lu2022}. Varying the Gilbert damping in our HC3TM model results in curves that deviate slightly from those presented in Fig. \ref{fig:magn_fluence} (data not shown) but the overall behaviour still remains the same.

To demonstrate the impact of lattice damping on the trend of the demagnetization more in detail, we have performed spin-lattice simulations for all three elements using several different values of the lattice damping. In addition to the simulations for the same lattice damping value ($\nu =1\times 10^{-13}$ kg/s - solid lines in Fig.\,\ref{fig:magn_fluence}), we also used lattice damping values corresponding to the upper and lower limits of the electron-phonon coupling $G_{ep}$ found in the literature \cite{PhysRevResearch.3.023032, PhysRevB.106.174407, PhysRevResearch.4.013104}. These intervals for the reported electron-phonon couplings are shown as the shaded areas in Fig.\,\ref{fig:magn_fluence} where it can be noticed that the trend of largest demagnetization for nickel and smallest demagnetization for cobalt holds over almost the full interval of the shaded areas. For all three elements, the lower bounds of the shaded areas in Fig.\,\ref{fig:magn_fluence} correspond to the lowest lattice damping while the upper bounds indicate the result of the largest lattice damping of the considered values. This further exemplifies the general picture that a stronger electron-lattice coupling transfers more heat from the electron to the lattice sub-system thus effectively reducing the amount of heat that gets pumped into the electron system.

In Ref. \cite{scheid2023uncovering} Scheid et al. experimentally showed that the maximum demagnetization amplitude in Fe and Co are surprisingly similar. Various reasons behind the distribution of demagnetization rates in iron, cobalt, nickel were discussed, including the impact of electron-phonon coupling, the electron-magnon scattering rate, and the ultrafast light-induced quenching of the inter-atomic exchange. In our model, as seen Fig.\,\ref{fig:magn_fluence}. this similarity could be explained by a very large lattice damping in Fe relatively to Co. Indeed, in HC3TM, and by using first principle values of Gilbert, to obtain similar slopes for iron and cobalt, we need to use a lattice damping for iron 10 times larger than for cobalt. This is illustrated in Fig.\,\ref{fig:magn_fluence} where the dashed curves indicate a lattice damping of $\nu =5 \times 10^{-13}$ kg/s for iron and $\nu =5\times10^{-14}$ kg/s for cobalt. 
However recent experimental \cite{PhysRevResearch.4.013104} and first principles \cite{PhysRevB.102.214305} results suggest that the lattice damping in cobalt is in fact larger than in iron. This would amplify the demagnetization in iron, while minimizing the one of cobalt. 
That conclusion of Scheid et al. \cite{scheid2023uncovering} is therefore in line with our findings, because even for the same lattice damping for all systems (which in our model corresponds to electron-phonon coupling, see Fig.\,\ref{fig:TM_scheme}) we clearly obtain the highest demagnetization rate for nickel followed by iron and then cobalt (see solid lines in Fig.\,\ref{fig:magn_fluence}).
One may further improve the agreement with the experimental observations by considering additionally ultrafast light-induced quenching of the interatomic exchange as suggested in Ref.\,\cite{scheid2023uncovering}. 

\section{Conclusions}

Using a heat-conserving three-temperature model we have calculated spin, lattice, and electron temperatures in simulations of the ultrafast magnetization dynamics of cobalt and iron. We have studied in detail the impact of Gilbert and lattice damping and compared results of the magnetization dynamics for bcc Fe, fcc Co and fcc Ni. It was found that in studies of ultrafast magnetic phenomena, the lattice dynamics plays a surprisingly important role, even in cases when the direct spin-lattice coupling is not significant (i.e., when Eq.\,\ref{SLHAM} is neglected). Simulations for various laser fluences were performed to study and compare ultrafast demagnetization of iron, cobalt, and nickel in various regimes. We have demonstrated that the simulations are consistent with recent observations, which show a linear trend between the maximally demagnetized state and laser fluence. We furthermore show that the experimentally found trend that for the same fluence, fcc Ni demagnetizes the most and fcc Co the least, holds for a wide range of realistic choices of the electron-phonon coupling strength. 

\section{ACKNOWLEDGEMENT}

This work was financially supported by the Knut and Alice Wallenberg Foundation (grant numbers 2018.0060, 2021.0246, and 2022.0108), 
Vetenskapsrådet (grant numbers 2019-03666, 2016-05980, and 2019-05304), 
the European Research Council (grant number 854843-FASTCORR), the foundation for Strategic Research SSF, and Olle Engkvist foundation.
Support from STandUP and eSSENCE is also acknowledged.
Computations were enabled by resources provided by the Swedish National Infrastructure for Computing (SNIC) at NSC, partially funded by the Swedish Research Council through grant agreement no. 2018-05973. P.S. acknowledges support from the ANR-20-CE09-0013
UFO. 

\appendix

\section{Heat capacities of the spin, lattice, and electron subsystems}
\label{AppendixA}
\begin{figure}[htb]
\includegraphics[width=0.47\textwidth]{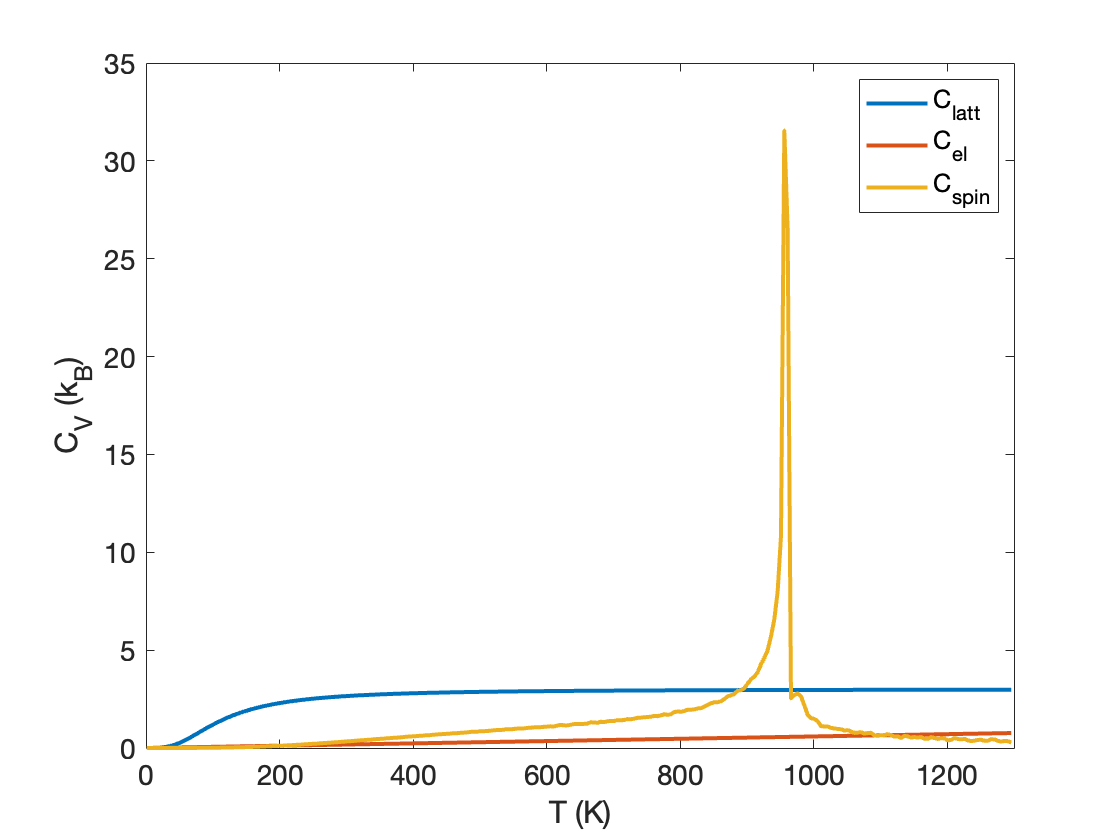}
\caption{\label{fig:Q_capacity} Heat capacities for the electronic, spin, and lattice subsystems used in the calculations for bcc Fe.}
\end{figure}

In our calculations we use temperature-dependent heat capacities for the lattice, electron, and spin subsystems. The corresponding curves are presented in Fig.\,\ref{fig:Q_capacity}.

The lattice heat capacity is obtained from the Debye model. The spin heat capacity is calculated using a recently proposed approach from quantum statistics \cite{Bergqvist2018}. The method more accurately describes heat capacities at low temperatures than what is obtained from Boltzmann statistics. However, its disadvantage is an overestimation of the heat capacity around $T_c$, as can be seen from Fig.\,\ref{fig:Q_capacity}.  Since in our calculations the spin temperature does not reach $T_c$, this shortcoming of the method does not impact our results. Moreover, our calculations show that using this approach does not change any of our main conclusions or results. We can confirm that all results and conclusions such as linear dependence of demagnetization amplitude on fluence, etc remain the same regardless of heat capacities used in calculations (please see Appendix A in \cite{PhysRevB.106.174407} for details of heat capacities calculations), including constant (i.e., temperature independent) heat capacities. The only thing that is affected by the change of capacities calculation is demagnetization amplitude. For the electronic subsystem, we assume that the electronic heat capacity is proportional to the electronic temperature $C_e = \gamma_e T_e$, and for bcc Fe we use $\gamma_e = \unit[4.9 \cdot 10^{-3}]{J\:mol^{-1}\:K^{-2}}$.

\section{Details of the \textit{ab-initio} calculations}

Similarly to Ref.\,\cite{PhysRevB.106.174407}, the calculation of the interactions to parameterize Eqs.\,\ref{eq1} and \ref{eq3} were performed in the framework of Density Functional Theory (DFT), using two different methods: (\textit{i}) the real-space linear muffin-tin orbital method in the atomic sphere approximation (RS-LMTO-ASA) \cite{Peduto1991,Frota-Pessoa1992} and (\textit{ii}) the plane-wave pseudopotential-based Quantum ESPRESSO (QE) package \cite{Giannozzi2017} in combination with the PHONOPY \cite{Togo2015} software. On one hand, this choice is motivated, essentially, on the well-known shortcoming of the LMTO-ASA method on obtaining accurate Hellmann-Feynman forces \cite{Wills2010} -- and, by consequence, phonon frequencies. On the other hand, to calculate the spin-lattice couplings (SLC), a supercell approach is necessary because of the breaking of inversion symmetry and long-range interactions \cite{PhysRevB.106.174407}. In this sense, RS-LMTO-ASA is more suitable for this calculation because it deals directly with real-space. 

For the computation of the force constants using a combination of QE and PHONOPY, the fcc Ni case used the same definitions as described in Ref.\,\cite{PhysRevB.106.174407}. Analogously, in the case of bcc Fe and fcc Co, in QE the scalar relativistic schemes based on the projector augmented-wave method (PAW) \cite{Bloechl1994} and the ultrasoft pseudopotentials (USPP) \cite{Vanderbilt1990}, respectively, are employed. As the exchange-correlation (XC) functional, we use the generalized gradient approximation (GGA) with the Perdew-Burke-Ernzerhof parametrization \cite{Perdew1996}. Here, the choice of a spin-polarized GGA XC term is motivated by the fact that the calculated lattice constants as well as the phonon frequencies better reproduce the experimental data in comparison with the local spin density approximation \cite{DalCorso2000}. The iron and cobalt atoms were described by standard pseudopotentials from the QE library, with $3s$, $3p$, $4s$, $4p$ and $3d$ valence electrons. A cutoff of 100 (1080) and 90 (500) Ry for the kinetic energy (charge density) was considered for fcc Co and bcc Fe, in this order. In turn, a Monkhorst-Pack (MP) \cite{Monkhorst1976} grid of $24\times24\times24$ $\vec{k}$-points was set for the first Brillouin zone integration. The self-consistent calculations with a $10^{-10}$ Ry convergence threshold were carried out using the Marzari-Vanderbilt \cite{Marzari1999} cold smearing with a spreading of 0.01 Ry. The force constants and phonon frequencies, computed by PHONOPY, were based on $6\times6\times6$ supercells (216 atoms), where we considered a $4\times4\times4$ $\vec{k}$-points mesh in the self-consistent cycle.

The exchange parameters, Gilbert damping ($\alpha$), density of states and spin-lattice couplings (SLC) were obtained using the RS-LMTO-ASA method. In this real-space formalism, the eigenvalue problem is solved with the help of the recursion method \cite{Haydock1980}, where the recursion chain is ended after $LL$ steps by making use of the Beer-Pettifor terminator \cite{Beer1984}. For bcc Fe we considered $LL = 31$, while for fcc Co a much higher $LL$ value ($LL=51$) is needed to better describe the density of states and Green’s functions at the Fermi level. In line with recent experience on finding a model to calculate reliable $\alpha$'s from first principles \cite{Lu2022}, the bulk systems consist of a big cluster in real-space containing $\sim55000$ (bcc) and $\sim696000$ (fcc) atoms, located in the perfect crystal positions. Differently from the QE calculations (in which the structures were optimized), the lattice parameters were fixed to the experimental values of $a=2.87\,$\AA$\,$(bcc Fe) and $a=3.54\,$\AA$\,$(fcc Co) \cite{American1972}. In this setup, the local density approximation (LSDA), with the von Barth and Hedin parametrization \cite{Barth1972}, was used. In particular, this choice of XC was based on the fact that some of quantities that parameterize Eqs.\,\ref{eq1}-\ref{SLHAM} were already investigated using LSDA \cite{Lu2022,Pajda2001,Yaroslav2015,Szilva2022}. It is a known trait, however, that GGA produces quantities with similar quality for the $3d$ magnetic elements \cite{Szilva2022,Grechnev2007,Lu2022}. Finally, the spin-orbit coupling (SOC) is included as a $l\cdot s$ term, computed in each variational step \cite{Andersen1975,Frota-Pessoa2004}. 

\label{AppendixB}

\bibliographystyle{apsrev4-2}
\bibliography{apssamp}


\end{document}